\documentclass[aps,superscriptaddress,nofootinbib,eqsecnum,prd,notitlepage,twocolumn, showkeys]{revtex4-1} 

\pdfoutput=1

\usepackage{amsfonts}
\usepackage{amsmath}
\usepackage{amssymb}
\usepackage{graphicx,color}
\usepackage{float}
\usepackage{hyperref}
\usepackage{subfigure}
\usepackage{dcolumn}
\usepackage{soul}
\usepackage{ulem}
\usepackage{verbatim}

\begin{document}

\title{Running and running of the running of the scalar spectral index in warm inflation}

\author{Suratna Das}
\email{suratna.das@ashoka.edu.in}
\affiliation{Department of Physics, Ashoka University,
   Rajiv Gandhi Education City, Rai, Sonipat: 131029, Haryana, India}

\author{Rudnei O. Ramos} \email{rudnei@uerj.br}
\affiliation{Departamento de Fisica Teorica, Universidade do Estado do
  Rio de Janeiro, 20550-013 Rio de Janeiro, RJ, Brazil }
\affiliation{Physics Department, McGill University, Montreal, Quebec, H3A 2T8, Canada}

\begin{abstract}

Next generation of cosmological observations are expected to improve
the measurements of several quantities connected to the primordial
inflation in the early Universe. These quantities include for example
improved measurements for the spectral index of the scalar 
curvature of the primordial power spectrum and to also bring a better
understanding on the scaling dependence of the primordial
spectrum. This  includes the running of the tilt and, possibly, also
the running of the running.  In this paper, we investigate the
possibility of generating large runnings in the context of warm
inflation.  Useful analytical expressions for the runnings are derived
in the context of warm inflation in the large dissipation regime. The
results are compared to and discussed for some well motivated
primordial inflaton potentials that have recently been of interest in
the literature.


\end{abstract}

\maketitle

\section{Introduction}

The final release of the {\it Planck} Cosmic Microwave Background
(CMB) data~\cite{Planck:2018jri} has severely constrained many
primordial observables, such as the scalar spectral index, $n_s$, the
tensor-to-scalar ratio, $r$, the primordial non-Gaussianities and the
primordial isocurvature spectrum, leaving us with the simplest vanilla
inflationary models as the most preferred ones. However, the presence
of the running, $\alpha_s$, and the running of running, $\beta_s$, of
the scalar spectral index that were inferred by the {\it Planck} data
may hint otherwise. {}For the $\Lambda$CDM model, the {\it Planck}
2018 TT(TT,TE,EE)+lowE+lensing data constrains the scalar spectral
index, its running and the running of its running
as~\cite{Planck:2018jri}
\begin{eqnarray}
  n_s&=&0.9587\pm0.0056\,\,(0.9625\pm0.0048)\nonumber\\
  \alpha_s&=&0.013\pm0.012\,\,(0.002\pm0.010)\nonumber\\
  \beta_s&=&0.022\pm0.012\,\,(0.010\pm0.013),
\end{eqnarray}
all at $68\%$ CL. According to {\it Planck}, such values for the
running and running of running of the scalar spectral index yield a
better fit to the low-$\ell$ deficit of the TT spectrum. It is
interesting to note that in these constraints there is a slight
preference for a positive and rather large $\beta_s\sim10^{-2}$, which
is also larger than the running of the spectral index
($\beta_s>\alpha_s$).  This is unexpected, when compared, for
instance, to the values derived from standard inflationary
models~\cite{Munoz:2016owz}, where it is expected an hierarchy like
$n_s \gg \alpha_s \gg \beta_s$. It has been shown in
Refs.~\cite{Garcia-Bellido:2014gna,Escudero:2015wba} that if we
consider the empirical relation $n_s-1\propto 1/N$, where $N$ is the
number of $e-$foldings, then in these vanilla models one expects 
\begin{eqnarray}
|\alpha_s|\sim\frac{1}{N^2}\lesssim 10^{-4}, \quad \quad
|\beta_s|\sim\frac{1}{N^3}\lesssim 10^{-5}.
\end{eqnarray}
The {\it Planck}--preferred vanilla inflationary models in general
also produce negative $\alpha_s$ and $\beta_s$.   Therefore, such
inflation models have been considered to be in tension with the
current observations by {\it
  Planck}~\cite{Escudero:2015wba,vandeBruck:2016rfv}. Going beyond
these vanilla models does not seem to improve the scenario either. It
was shown in Ref.~\cite{vandeBruck:2016rfv} that single field
inflation models and non-interacting two-field models are incapable of
producing a $\beta_s$ larger than $\alpha_s$. Slow-roll violating
models and those with non-trivial evolution of sound speed could be
able to provide exception to this trend, but with a considerable
amount of fine-tuning~\cite{vandeBruck:2016rfv}. 

The {\it Planck}--preferred vanilla inflaton potential models in
general consider the traditional cold inflation (CI) scenario, where
the dynamics of the inflaton field $\phi$ is assumed to be independent
of its couplings with other fields during inflation. However, such
interaction terms play a major role at the end of inflation, when the
inflaton field needs to release its energy density in the form of
radiation through decay processes to reheat the Universe and, hence,
leading to a radiation dominated Universe as required by the big bang
cosmology. On the other hand, in the warm inflation (WI)
scenario~\cite{Berera:1995ie}, the couplings between the inflaton and
other fields are strong enough such that their effects on the inflaton
dynamics cannot be ignored. In WI (for reviews on WI, see, e.g.,
Refs.~\cite{Berera:2008ar,BasteroGil:2009ec}), the inflaton field is
able to keep dissipating its energy such that a non-negligible
radiation bath throughout inflation can be produced, while preserving
the flatness of the inflaton potential required for slow-rolling of
the inflaton field. The presence of such a quasi-equilibrium thermal
radiation bath during WI helps transiting smoothly from the
inflationary accelerated phase to the radiation dominated
phase after inflation ends, avoiding a phase of reheating in
between. In WI, the inflaton dynamics are modified with respect to the
one in CI due to the presence of an extra friction term,
$\Upsilon\dot\phi$, that accounts for the energy transfer between the
inflaton field and the radiation bath present during inflation. Due to
such modified dynamics, the inflationary observables, like $r$, $n_s$
and the non-Gaussianity parameter $f_{\rm NL}$, are also modified with
respect to those obtained in CI. These changes have certain
advantages. {}For example, some of the inflaton potentials excluded by
data in the context of CI, can be made in tune with the observations
in the WI context. One such example being the quartic chaotic
potential $\lambda\phi^4$ \cite{Bartrum:2013fia}. Besides, it has
recently been shown that, while CI fails to comply with the recently
proposed Swampland Conjectures in String 
Theory~\cite{Ooguri:2018wrx,Garg:2018reu,Kinney:2018nny}, 
WI can easily accommodate those
criteria~\cite{Das:2018hqy,Motaharfar:2018zyb,Das:2018rpg,Das:2019hto,Berera:2019zdd,Berera:2020dvn}. 
Hence, WI provides a way to construct inflationary
models that can be consistent as effective models that could descent
from an ultraviolet complete quantum gravity, despite the Swampland
Conjectures barring them from constructing de Sitter vacua in String
Landscapes. 

The running and the running of the running of the scalar spectral
index in the context of WI were first studied in
Ref.~\cite{Benetti:2016jhf}, where several inflationary potentials
were analyzed with two different forms of dissipative terms considered
in WI, $\Upsilon_{\rm cubic}\propto T^3/\phi^2$ and $\Upsilon_{\rm
  linear}\propto T$ ($T$ being the temperature of the thermal
bath). The WI models studied in Ref.~\cite{Benetti:2016jhf} were,
however, treated in the weak dissipative regime ($Q\ll1$, where Q is
the ratio of the thermal friction term $\Upsilon\dot\phi$ and the
Hubble expansion friction term $3H\dot\phi$ present in the inflaton
equation of motion in WI). This was because the models studied in that
reference could only lead to consistent observables, e.g, values for
$r$ and $n_s$, in that specific dissipation regime of WI. In
particular, it was shown in Ref.~\cite{Benetti:2016jhf} that in all
the models studied there, in the weak dissipative regime, there was
still a large hierarchy between the values of $\alpha_s$  and
$\beta_s$. 

The aim of this paper is to study the running and the running of the
running of the scalar spectral index in the context of WI, where WI is
realized in a strong dissipative regime $(Q\gg1)$. This is motivated
by the recent results in WI in the context of the swampland
conjectures~\cite{Das:2018hqy, Motaharfar:2018zyb, Das:2018rpg,
  Das:2019hto} and also on the solution of the so-called
$\eta$-problem~\cite{Bastero-Gil:2019gao}, which exactly favors WI
being realized in the strong dissipative regime. {}For this purpose,
we will focus mostly in a certain WI model, dubbed the Minimal Warm
Inflation (MWI) \cite{Berghaus:2019whh}, where WI has been shown to be
possible to be realized in the strong dissipative regime and in which
the dissipative term is proportional to the cubic power of the
temperature of the thermal bath $(\Upsilon\propto T^3)$. However, our
results will be kept as general as possible, such that they can be
extended to other models. In the original paper of
MWI~\cite{Berghaus:2019whh}, a hybrid potential was used to produce a
red-tilted scalar power spectrum. Later, MWI was studied with
generalized exponential potentials in Ref.~\cite{Das:2020xmh}, and it
was shown that this model is not only in accordance with the current
observations (yielding the appropriate values for $n_s$ and $r$), but
also are in tune with the Swampland Conjectures. We will analyze both
the original MWI model, with the hybrid potential, and also the  one
with the generalized exponential potential, to derive $\alpha_s$ and
$\beta_s$ that are produced in these two scenarios. It is to be
noticed that although the running and the running of the running of
the scalar spectral index in the context of WI were first studied in
Ref.~\cite{Benetti:2016jhf}, they were estimated numerically. Here,
however, we aim to produce explicit analytical expressions for these
quantities in WI. To the best of our knowledge, this is  the first
time that such analytical analysis and derivation of $\alpha_s$ and
$\beta_s$ in the context of WI are presented. Given the advent of new
generations of cosmological observatories probing both the cosmic
microwave background (CMB)
measurements~\cite{CMB-S4:2022ght,Chang:2022tzj}, the distribution of
matter at low-redshift from optical, near-infrared, and 21cm intensity
surveys~\cite{Pritchard:2011xb}, it is expected  that theories of
cosmic inflation will be further constrained with the more precise
cosmological data. Thus, it is important to have such analytical
expressions to help in our ability in finding possible models of
interest and also in model building in WI.

The remainder of the paper is organized as follows. In
Sec.~\ref{general}, we briefly review WI and present the general
expressions for $n_s$, $\alpha_s$ and $\beta_s$. In Sec.~\ref{MWI}, we
analyze  the case of MWI with a hybrid potential. In
Sec.~\ref{MWI-expo}, we turn our attention to the determination of
$n_s$, $\alpha_s$ and $\beta_s$ in the recently studied generalized
exponential potentials for the inflaton in WI. Then, in
Sec.~\ref{conclusion}, we discuss our findings and conclude. 

\section{Running and running of the running of scalar spectral index in WI - general expressions}
\label{general}

The equation of motion of the inflaton field $\phi$ in WI is
\begin{eqnarray}
\ddot{\phi}+3H(1+Q)\dot\phi+V,_\phi=0,
\label{eom}
\end{eqnarray}
where $Q$ is the ratio of the two frictional terms, the friction due
thermal bath and the friction due to Hubble expansion, present in the
theory,
\begin{eqnarray}
Q\equiv \frac{\Upsilon(T,\phi)}{3H},
\label{ratioQ}
\end{eqnarray}
where the dissipation coefficient $\Upsilon\equiv \Upsilon(T,\phi)$ is
in general a function of the temperature of the thermal radiation bath
generated and can also depend on the inflaton amplitude (for examples
of different forms of dissipation coefficients derived in quantum
field theory and used in WI, see, e.g.,
Ref.~\cite{BasteroGil:2010pb}).  In Eq.~(\ref{eom}),  $V,_\phi\equiv
dV/d\phi$, where $V$ is the potential of the inflaton field. As the
inflaton dissipates part of its energy density, it can sustain a
radiation bath, with energy density $\rho_r$ and whose evolution
equation is given by
\begin{eqnarray}
\dot \rho_r + 4 H \rho_r = 3 H Q \dot \phi^2.
\label{eomrhor}
\end{eqnarray}
We define the slow-roll parameters $\epsilon_V$ and $\eta_V$ in the
usual way,
\begin{eqnarray}
\epsilon_V&=&\frac{M_{\rm Pl}^2}{2}\left(\frac{V,_\phi}{V}\right)^2,
\\ \eta_V&=&M_{\rm Pl}^2\frac{V,_{\phi\phi}}{V},
\label{slowrollV}
\end{eqnarray}
where $M_{\rm Pl} \equiv 1/\sqrt{8 \pi G} \simeq 2.4 \times
10^{18}$GeV is the reduced Planck mass.   In WI, another set of
slow-roll parameters are also defined,
\begin{eqnarray}
\epsilon_{\rm WI}&=&\frac{\epsilon_V}{1+Q},\\ \eta_{\rm
  WI}&=&\frac{\eta_V}{1+Q},
\label{slowrollWI}
\end{eqnarray}
due to the fact that WI actually ends when $\epsilon_{\rm WI}\sim1$,
or, similarly, when $\epsilon_V\sim 1+Q$, while in CI the usual
condition for the end of the accelerated inflationary regime is simply
$\epsilon_V \sim 1$. 

Besides $\epsilon_V$ and $\eta_V$, it is also useful to define the
higher order slow-roll coefficients $\xi^2_V$ and $\omega^3_V$,
as~\cite{Zarei:2014bta}
\begin{eqnarray}
\xi_V^2\equiv M_{\rm Pl}^4\frac{V,_\phi V,_{\phi\phi\phi}}{V^2},
\label{xi2V}
\end{eqnarray}
and
\begin{eqnarray}
\omega_V^3=M_{\rm Pl}^6\frac{V,_\phi^2V,_{\phi\phi\phi\phi}}{V^3},
\end{eqnarray}
which will appear in the equations for the running and for the running
of the running to be derived later on.

The primordial scalar curvature power spectrum of a typical WI
model can be written as~\cite{Ramos:2013nsa,Benetti:2016jhf}
\begin{eqnarray}
\Delta_{\mathcal R}(k/k_*)=\left(\frac{H_*^2}{2\pi\dot\phi_*}\right)^2\mathcal{F}(k/k_*),
\label{full-power}
\end{eqnarray}
where the subindex $*$ means that the quantities are evaluated at the
Hubble radius crossing of the pivot scale $k_*$.  Here, the function
$\mathcal{F}(k/k_*)$ is given as 
\begin{eqnarray}
\mathcal{F}(k/k_*)\equiv\left(1+2n_*+\frac{2\sqrt3\pi
  Q_*}{\sqrt{3+4\pi Q_*}}\frac{T_*}{H_*}\right)G(Q_*), \nonumber \\
\label{Fk}
\end{eqnarray}
where $n_*$ is the thermal distribution of the inflaton field due to
the presence of the radiation bath and $G(Q_*)$ accounts for the
effect of the coupling of the inflaton and radiation fluctuations
\cite{Graham:2009bf, Bastero-Gil:2011rva, Bastero-Gil:2014jsa}. 

The primordial scalar power spectrum $\Delta_{\mathcal R}(k/k_*)$ can
be expanded in terms of the small scale-dependence as
\begin{eqnarray}
\Delta_{\mathcal R}(k/k_*) \simeq  A_s
\left(\frac{k}{k_*}\right)^{n_s-1 + \frac{\alpha_s}{2} \ln(k/k_*)
  +\frac{\beta_s}{6} \ln^2(k/k_*)},
\label{DeltaR}
\end{eqnarray}
where $A_s$ is the scalar amplitude, $n_s$ is the scalar tilt,
$\alpha_s$ is the running and $\beta_s$ is the running of the running
(or second running).

{}From Eq.~(\ref{DeltaR}), the scalar spectral index $n_s$ can be
calculated at the horizon crossing ($k=k_*$) as
\begin{equation}
n_s -1 =   \frac{d \ln \Delta_{{\cal R}}(k/k_*) }{d \ln(k/k_*)}\Bigr|_{k\to k_*} .
\label{eq:n}
\end{equation}
Likewise, the expressions for $\alpha_s$ and $\beta_s$ are determined,
respectively, by
\begin{equation}
\alpha_s=   \frac{dn_s(k/k_*) }{d \ln(k/k_*) }\Bigr|_{k\to k_*},
\label{dns}
\end{equation}
and
\begin{equation}
\beta_s= \frac{d^2 n_s(k/k_*) }{d \ln(k/k_*)^2 }\Bigr|_{k\to k_*} =
\frac{d \alpha_s(k/k_*) }{d \ln(k/k_*) }\Bigr|_{k\to k_*}.
\label{d2ns}
\end{equation}

Using that at Hubble radius crossing $k_*=aH$, we can write $d\ln
k=(d\ln k/dN) dN$, with $N=\ln a$ being the number of $e$-folds and
$\epsilon_H=-\dot H/H^2$, then
\begin{equation}
\frac{d\ln k}{dN}=1-\epsilon_H\approx  1-\epsilon_V/(1+Q).
\label{dlnk}
\end{equation}

The definitions given by Eqs.~(\ref{eq:n}), (\ref{dns}) and
(\ref{d2ns}) can now be applied directly to the scalar curvature in
WI, Eq.~(\ref{full-power}). Once an appropriate functional form for
the dissipation coefficient is given, explicit expressions for $n_s$,
$\alpha_s$ and $\beta_s$ can be derived. Here we work with the
well-motivated functional form for the dissipation
coefficient\footnote{{}For earlier studies considering this functional
  form for the dissipation coefficient in WI, see, e.g.,
  Refs.~\cite{Zhang:2009ge,Visinelli:2016rhn}.}
\begin{equation}
\Upsilon(\phi,T)=C_\Upsilon\,T^p \phi^c {\cal M}^{1-p-c},
\label{Upsilon}
\end{equation}
where $C_\Upsilon$ is a constant, $T$ is the temperature and ${\cal M}$ is
some appropriate mass scale. {}For specific examples of microscopic
quantum field theory derivations of such dissipation coefficients in
WI,  see, e.g.,
Refs.~\cite{Gleiser:1993ea,Berera:1998gx,Berera:2008ar,BasteroGil:2012cm,Bastero-Gil:2016qru}.

When taking derivatives of Eq.~(\ref{full-power}) such as to obtain $n_s$, $\alpha_s$ and $\beta_s$,
we are naturally faced with accounting for derivatives on $Q$ and $T/H$ with respect to $N$.
These expressions depend on the specific form of the dissipation coefficient used.
{}For completeness, let us quote them here for the generic dissipation term Eq.~(\ref{Upsilon}).
We use the slow-roll approximation for the background dynamical equations for the inflaton field and 
for the radiation energy density in WI, e.g.,
\begin{eqnarray}
&&3H(1+Q)\dot\phi\approx -V,_{\phi},
\label{sr1}\\ 
&&\rho_R\equiv
  C_RT^4\approx\frac{3Q}{4}\dot\phi^2,
\label{sr2}
\\ 
&&H^2\approx \frac{V}{3M_{\rm Pl}^2}.
\label{sr3}
\end{eqnarray}
{}From these set of equations, one can easily deduce for instance that
\begin{eqnarray}
Q^7\propto \frac{V,_\phi^6}{V^5},\quad\quad T^7\propto
\frac{V,_\phi^2}{V^{1/2}}.
\end{eqnarray}
Thus, from Eqs.~(\ref{sr1}), (\ref{sr2}) and (\ref{sr3}), together with Eq.~(\ref{Upsilon}),
we obtain that
\begin{eqnarray}
\frac{d \ln Q}{dN} &=& \frac{2 Q \left[(2+p) \epsilon_V -p \eta_V -2 c \kappa_V \right]}{4-p+(4+p) Q},
\label{dlnQdN}
\end{eqnarray}
and
\begin{eqnarray}
\frac{d \ln \left(\frac{T}{H}\right)}{dN} &=&
\frac{ \left[7- (1-Q)p+5 Q\right] \epsilon_V  }{(1+Q) \left[4-p+(4+p) Q\right]}
\nonumber \\
&-& \frac{ 2 (1+Q) \eta_V + (1-Q)c\kappa_V }{(1+Q) \left[4-p+(4+p) Q\right]},
\label{dlnTHdN}
\end{eqnarray}
where we have also introduced the quantity $\kappa_V$, defined as
$\kappa_{V} = M_{\rm Pl}^2 V_{,\phi}/(\phi \,V)$.
{}From Eqs.~(\ref{dlnQdN}) and (\ref{dlnTHdN}), we find in particular that
\begin{eqnarray}
n_s&=& 1+\frac{(1+Q)}{1+Q-\epsilon_V} \left[\frac{d \ln \left(\frac{T}{H}\right)}{dN}
\right.
\nonumber\\
&+& \left. \frac{\frac{d \ln Q}{dN} 
\left(-3+Q \left\{3+2 \pi  \left[-1+Q \left(3+\sqrt{9+12 Q \pi
}\right)\right]\right\}\right)}{(1+Q) \left[3+Q \pi  \left(4+\sqrt{9+12 Q \pi }\right)\right]}
\right.
\nonumber \\
&+& \left.\frac{d \ln Q}{dN} {\cal A}(Q)+\frac{-6
\epsilon_V+2 \eta_V}{1+Q}\right],
\label{generalns}
\end{eqnarray}
where we have defined ${\cal A}(Q)$ as
\begin{eqnarray}
\mathcal{A}(Q)=\frac{3+2\pi Q}{3+4\pi Q}+Q\frac{d\ln G(Q)}{dQ}.
\label{AQ}
\end{eqnarray}

The expressions for $\alpha_s$ and $\beta_s$ are obtained by taking further
derivatives of Eq.~(\ref{generalns}) using Eq.~(\ref{dlnk}). Though this is a
lengthy but straightforward exercise, the  expressions are, 
however, too long to quote here in full. But since we are mostly interested
in models of WI in the strong dissipative regime, we can expand 
Eq.~(\ref{generalns}) for $Q\gg 1$ and obtain, at least for $n_s$, a relatively shorter expression.
Hence, 
expanding Eq.~(\ref{generalns}) for  $Q\gg 1$, we obtain that
\begin{widetext}
\begin{eqnarray}
n_s &\simeq & 1+\frac{-7 c \kappa_V-11 \epsilon_V-p \epsilon_V+6 \eta_V-2 p \eta_V}{Q (4+p)}
+ \frac{2 {\cal A}(Q) \left(-2 c \kappa_V+(2+p) \
\epsilon_V-p \eta_V \right)}{Q (4+p)} 
+ {\cal O}(1/Q^{\frac{3}{2}}),
\label{nslargeQ}
\end{eqnarray}
where we have used also Eqs.~(\ref{dlnQdN}) and (\ref{dlnTHdN}).
Approximate expressions  for $\alpha_s$ and $\beta_s$ can be obtained analogously.
{}For instance, from Eq.~(\ref{nslargeQ}) we obtain for $\alpha_s$ that
\begin{eqnarray}
\alpha_s &\simeq& \frac{1}{(4+p)^3 Q^2} 
\left\{
-4 (4+p) {\cal A}(Q) \left(2 c \kappa_V-(2+p) \epsilon_V+p \eta_V \right){}^2
+ 2 (4+p)^2 {\cal A}(Q) \left(4 (2+p) \epsilon_V^2-4 (1+p) \epsilon_V \eta_V+p \xi_V^2\right)
\right.
\nonumber \\
&+& \left.  (4+p) \left(-2 (6+p) (11+p) \epsilon_V^2
+ 2 \epsilon_V \left(c (-8+5 p) \kappa_V+56 \eta_V \right)-2 \left(7 c \kappa_V
+2 (-3+p) \eta_V \right) \left(2 c \kappa_V+p \eta_V \right)
\right.\right.
\nonumber \\
&-& \left.\left. 24 \xi_V^2+2 p (1+p) \xi_V^2\right)+4 \left(4 \left(-1+Q+\epsilon_V \right)+p
\left(1+Q+\epsilon_V \right)\right) \left(2 c \kappa_V-(2+p) \epsilon_V+p \eta_V \right){}^2 {\cal A}'(Q)
\right\} + {\cal O}\left(\frac{1}{Q^3}\right),
\nonumber \\
\label{alphaslargeQ}
\end{eqnarray}
while for $\beta_s$ we find
\begin{eqnarray}
\beta_s &\simeq &
\frac{2}{(4+p)^3 Q^3} 
\left\{
8 {\cal A}(Q) \left(-2 c \kappa_V+(2+p) \epsilon_V-p \eta_V \right){}^3
-8 (4+p) {\cal A}(Q) \left(-2 c \kappa_V+(2+p) \epsilon_V-p \eta_V \right)
\left(4 (2+p) \epsilon_V^2
\right.\right.
\nonumber \\
&-& \left. \left. 4 (1+p) \epsilon_V \eta_V+p \xi_V^2\right)-2 \left(-2 c \kappa_V+(2+p) \epsilon_V
-p \eta_V \right) \left(-2 (6+p)(11+p) \epsilon_V^2+2 \epsilon_V \left(c (-8+5 p) \kappa_V+56 \eta_V \right)
\right. \right.
\nonumber \\
&-& \left. \left. 2 \left(7 c \kappa_V+2 (-3+p) \eta_V \right) \left(2 c \kappa_V+p \eta_V \right)
-24 \xi_V^2+2 p (1+p) \xi_V^2\right)+(-4-p) \left(8 (6+p) (11+p) \epsilon_V^3
\right. \right. 
\nonumber \\
&+& \left. \left. 8 \epsilon_V \eta_V \left(c (-5+4 p) \kappa_V+(14+(-3+p)
p) \eta_V \right)-4 \epsilon_V^2 \left(c (-8+5 p) \kappa_V+(150+p (17+p)) \eta_V \right)
\right. \right.
\nonumber \\
&+& \left. \left. c (12-11 p) \kappa_V \xi_V^2-4 \left(-26+p+p^2\right) \epsilon_V \xi_V^2
-(-3+p) (-4+3 p) \eta_V \xi_V^2+(-3+p) (4+p) \omega_V^3\right)
\right. 
\nonumber \\
&+& \left.  (4+p)^2 {\cal A}(Q) \left(32 (2+p) \epsilon_V^3-8 (7+5 p) \
\epsilon_V^2 \eta_V+4 \epsilon_V \left(2 (1+p) \eta_V^2+\xi_V^2+2 p \xi_V^2\right)
-p \left(\eta_V \xi_V^2+\omega_V^3\right)\right)
\right.
\nonumber \\
&-& \left. 4 Q \left(-2 c \kappa_V+(2+p) \epsilon_V-p \eta_V \right){}^3 {\cal A}'(Q)
\right.
\nonumber \\
&+& \left. 2 Q (4+p) \left(-2 c \kappa_V+(2+p) \epsilon_V-p \eta_V \right) \left(4 (2+p)
\epsilon_V^2-4 (1+p) \epsilon_V \eta_V+p \xi_V^2\right) {\cal A}'(Q)
\right.
\nonumber \\
&+& \left. 4 Q \left(-2 c \kappa_V+(2+p) \epsilon_V-p \eta_V \right) \left(-\left(2
c \kappa_V-(2+p) \epsilon_V+p \eta_V \right){}^2+(4+p) \left(4 (2+p) \epsilon_V^2
-4 (1+p) \epsilon_V \eta_V+p \xi_V^2\right)\right) {\cal A}'(Q)
\right.
\nonumber \\
&+& \left. 4 Q^2 \left(-2 c \kappa_V+(2+p) \epsilon_V-p \eta_V \right){}^3 {\cal A}''(Q)
\right\} + {\cal O}\left(\frac{1}{Q^4}\right).
\label{betaslargeQ}
\end{eqnarray}
\end{widetext}
In the above expressions, ${\cal A}'(Q)\equiv d{\cal A}/dQ$ and  ${\cal A}''(Q)\equiv d^2{\cal A}/dQ^2$.
The above approximate expressions for $n_s$, $\alpha_s$ and $\beta_s$ applies to any WI
model with a generic dissipation coefficient of the form of Eq.~(\ref{Upsilon}) and once
the primordial potential for the inflaton is specified.

Below, as an example, we will work with two explicit models in WI, which have been
shown to be of interest recently. Then, specific expressions for  $n_s$, $\alpha_s$ and $\beta_s$
will be derived for those models, along  with
their respective analysis.

\section{Running and running of the running of the scalar spectral index in MWI}
\label{MWI}

The minimal WI (MWI) model proposed in Ref.~\cite{Berghaus:2019whh}
(see also Ref.~\cite{Laine:2021ego}) successfully realizes WI in the
strong dissipative regime. In such a model, the inflaton is being
treated as an axionic field with coupling to the non-Abelian gauge
fields. Due to such gauge couplings, sphaleron transitions between
gauge vacua at high temperatures lead to a friction term of the form
\begin{eqnarray}
\Upsilon(T)=\frac{\Gamma_{sp}(T)}{2f^2T}=\kappa(\alpha_g, N_c,
N_f)\alpha_g^5\frac{T^3}{f^2},
\label{upsilon}
\end{eqnarray}
where $\Gamma_{sp}$ is the sphaleron rate, $f$ is the axion decay
rate, $T$ is the temperature, $\alpha_g=g^2/(4\pi)$, with $g$ being
the Yang-Mills gauge coupling, and $\kappa$ is a dimensionless
quantity depending on the dimension of the gauge group $(N_c)$, the
representation of the fermions $(N_f)$ and on the gauge coupling $g$
through $\alpha_g$. {}For instance, for a quantum chromodynamics (QCD)
type of axion and using typical values of parameters for QCD,
we have that $\kappa(\alpha_g, N_c,N_f) \alpha_g^5 \sim 10^{-3}$.

This model is particularly attractive in the
context of WI. Because of the axionic shift symmetry,  the inflaton
is protected from any perturbative backreaction and, hence, from
acquiring a large thermal mass. Similar symmetry properties allowing
the inflaton to be coupled directly to the radiation fields have also
been studied in
Refs.~\cite{Bastero-Gil:2016qru,Bastero-Gil:2019gao}. Models based on
pseudo-Goldstone bosons for the inflaton~\cite{Freese:1990rb} are
quite reminiscent of these ideas and have gained increased attention
recently in the context of WI~\cite{Montefalcone:2022jfw}. 

The scalar power spectrum in the MWI model that was considered in
Ref.~\cite{Berghaus:2019whh} was given by
\begin{eqnarray}
\Delta_{\mathcal R}\approx
\frac{\sqrt3}{4\pi^{\frac32}}\frac{H^3T}{\dot\phi^2}\left(\frac{Q}{Q_3}\right)^9Q^{\frac12},
\label{DeltaMWI}
\end{eqnarray}
where $Q_3\approx 7.3$. At the end of this section and in the next one, we will discuss a more accurate form for
the scalar spectrum in this model, for a dissipation coefficient of the form $\Upsilon \propto T^3$ as it is
been considered here. But for now, let us use the above exression Eq.~(\ref{DeltaMWI}) as
considered by the authors in Ref.~\cite{Berghaus:2019whh}.

Since we are also deriving expressions in the high dissipative regime, $Q \gg 1$, let us note that in this case,
using $p=3$ and $c=0$ in Eq.~(\ref{Upsilon}), that
\begin{eqnarray}
&&\frac{d\ln H}{dN}\approx-\frac{\epsilon_V}{Q},\\  &&\frac{d\ln
    Q}{dN}\approx  \frac{10\epsilon_V-6\eta_V}{7Q}, \\ && \frac{d\ln
    T}{dN}\approx
  \frac{\epsilon_V-2\eta_V}{7Q},\\  &&\frac{d\ln\dot\phi}{dN}
\approx-\frac{3\epsilon_V+\eta_V}{7Q}.
\label{equations}
\end{eqnarray}
Hence, 
the scalar spectral index, in the strong
dissipative regime can be calculated as 
\begin{eqnarray}
n_s-1=\frac{3}{7Q}(27\epsilon_V-19\eta_V).
\label{spectral-index}
\end{eqnarray}
{}For a detailed derivation of the scalar spectral index see, e.g.,
Ref.~\cite{Das:2020xmh}. 

It is evident from Eq.~(\ref{spectral-index}) that to obtain the
observed red tilt of the scalar spectral index $(n_s<1)$ in MWI, one
requires a potential which yields $\epsilon_V<\eta_V$. In
Ref.~\cite{Berghaus:2019whh}, for this objective it was considered a
potential of a hybrid inflation model. The effective potential of
hybrid inflation contains the inflaton field $\phi$ along with a
waterfall field $\sigma$, which can be written as\footnote{Note that
there are many variations of the hybrid inflation model, including
the effect of possible quantum correction terms (see, e.g., Ref.~\cite{Bastero-Gil:2004oun}).
In the present work, we only consider the potential for the hybrid model
in its traditional form, given by Eq.~(\ref{Vhybrid}), since our intention
is to compared our results with those from Ref.~\cite{Berghaus:2019whh}.}
\begin{eqnarray}
  V(\phi,\sigma)=\frac{1}{4\lambda}(M^2-\lambda\sigma)^2+\frac12m^2\phi^2
  +\frac12g^2\phi^2\sigma^2.\nonumber\\
\label{Vhybrid}
\end{eqnarray}
The $\sigma$ field has a squared mass term as $-M^2+g^2\phi^2$ in this
potential. When $\phi>M/g$, which is the region where inflation takes
place, $\sigma$ has only one minimum at $\sigma=0$, and the effective
potential becomes\footnote{In this work we are not concerned with 
possible issues related to formation
of topological defects that might result in  hybrid inflation associated 
with the symmetry breaking of the $\sigma$ field and which are related to the post-inflationary 
dynamics regime. {}For a discussion about topological defects in the context of
hybrid inflation, see, e.g., Ref.~\cite{Lyth:1998xn}.}
\begin{eqnarray}
V_{\rm eff}(\phi)=\frac{M^4}{4\lambda}+\frac12m^2\phi^2.
\label{pot}
\end{eqnarray}
During this stage, the constant term $\frac{M^4}{4\lambda}$ $(\gg
\frac12 m^2\phi^2 )$ drives the expansion.  Below the threshold
$\phi=M/g$, the waterfall field $\sigma$ quickly rolls down to its
minimum $\sigma(\phi)=M_\sigma(\phi)/\sqrt{\lambda}$ and puts an end
to inflation.   The effective potential, given in Eq.~(\ref{pot}),
yields the slow-roll parameters as\footnote{Note that in
  Ref.~\cite{Berghaus:2019whh}, $\epsilon_V$ was defined as
  $\epsilon_{\rm WI}$.}
\begin{eqnarray}
\epsilon_V&\approx&8\lambda^2\frac{M_{\rm
    Pl}^2m^4\phi^2}{M^8},\nonumber\\ \eta_V&\approx&4\lambda\frac{M_{\rm
    Pl}^2m^2}{M^4},
\label{slow-roll}
\end{eqnarray}
which shows that 
\begin{eqnarray}
\frac{\epsilon_V}{\eta_V}\sim\frac{\frac12m^2\phi^2}{\frac{M^4}{4\lambda}}\ll1.
\end{eqnarray}
Therefore, such a potential is suitable to produce a red-tilted scalar
spectral index in MWI. Hence, with the hybrid inflation potential
given in Eq.~(\ref{pot}), the scalar spectral index in
Eq.~(\ref{spectral-index}) can be approximated as 
\begin{eqnarray}
n_s-1\sim-\frac{57\eta_V}{7Q}.
\end{eqnarray}
Above all, as we can see from Eq.~(\ref{slow-roll}) that $\eta_V$ is
effectively constant in this model, we denote $\eta_V=k$, where
$k\equiv 4\lambda\frac{M_{\rm Pl}^2m^2}{M^4}$ is a constant. Hence, to
yield the observed value of the red-tilt, $1-n_s\sim0.04$, we require 
\begin{eqnarray}
\frac{k}{Q}\sim0.005.
\label{kQ}
\end{eqnarray}

We now estimate the running $(\alpha_s)$ and the running of the
running $(\beta_s)$ of the scalar spectral index in this model. The
running of the scalar spectral index, $\alpha_s$, can be calculated as 
\begin{eqnarray}
\alpha_s\simeq\frac{dn_s}{dN}=\frac{57}{7}\frac{k}{Q}\frac{d\ln
  Q}{dN},
\end{eqnarray}
where derivative has been taken with respect to the number of
$e-$foldings $N$, and in MWI we have~\cite{Das:2020xmh}
\begin{eqnarray}
\frac{d\ln Q}{dN}=\frac{10\epsilon_V-6\eta_V}{7Q}.
\end{eqnarray}
However, as the effective hybrid potential given in Eq.~(\ref{pot})
yields $\epsilon_V\ll\eta_V$, we can approximate the above equation as 
\begin{eqnarray}
\frac{d\ln Q}{dN}\approx-\frac{6k}{7Q}.
\end{eqnarray}
This yields the running of the scalar spectral index in MWI as 
\begin{eqnarray}
\alpha_s\approx-\frac{342}{49}\frac{k^2}{Q^2}.
\end{eqnarray}
Now, the value of $k/Q$, which yields the observed red-tilt of the
scalar spectral index given in Eq.(\ref{kQ}), produces a running of
the scalar spectral index as 
\begin{eqnarray}
\alpha_s\sim-1.7\times10^{-4}.
\label{alphas1}
\end{eqnarray}

The running of the running of the scalar spectral index then turns out
to be 
\begin{eqnarray}
\beta_s\simeq \frac{d\alpha_s}{dN}
=\frac{684}{49}\frac{k^2}{Q^2}\frac{d\ln
  Q}{dN}\approx-\frac{4104}{441}\frac{k^3}{Q^3}.
\end{eqnarray}
Again, the value of $k/Q$ given in Eq.~(\ref{kQ}), which yields the
observed red-tilt of the scalar spectral index, gives rise to a
running of the running of the scalar spectral index as 
\begin{eqnarray}
\beta_s\approx-1.16\times10^{-6}.
\label{betas1}
\end{eqnarray}
Therefore, we note that MWI model with hybrid-type potential generates
negative $\alpha_s$ and $\beta_s$ and the running of the running is
two orders of magnitude smaller than the running of the scalar
spectral index. These results come from the simplified analysis made
in particular with the approximated scalar power spectrum given by Eq.~(\ref{DeltaMWI}).

Let us now check whether an improved form for the scalar power spectrum might
somehow improve on the estimates made above for $n_s$, $\alpha_s$ and $\beta_s$.
Using the previously derived Eqs.~(\ref{nslargeQ}), (\ref{alphaslargeQ}) and 
(\ref{betaslargeQ}), we obtain for instance that
\begin{eqnarray}
n_s\simeq 1+\frac{2 {\cal A}(Q) \left(\frac{5 m^4 \phi ^2 M_{\text{Pl}}^2}{2 V_{\text{eff}}^2}-\frac{3 m^2 M_{\text{Pl}}^2}{V_{\text{eff}}}\right)}{7 Q}-\frac{m^4 \phi ^2 M_{\text{Pl}}^2}{Q V_{\text{eff}}^2},
\nonumber \\
\label{nsMWI}
\end{eqnarray}

\begin{widetext}
\begin{eqnarray}
\alpha_s &\simeq& \frac{1}{686 Q^2 V_{\text{eff}}^6} \left\{ 7 {\cal A}'(Q) m^8 M_{\text{Pl}}^6 \left(5 m^2 \phi ^3
-6 \phi  V_{\text{eff}}\right){}^2+2 m^4 M_{\text{Pl}}^4 V_{\text{eff}}^2 \left(m^4(-441+315 {\cal A}(Q)+
25 {\cal A}'(Q) (-1+7 Q)) \phi ^4
\right.\right.
\nonumber \\
&+& \left. \left. 4 V_{\text{eff}} \left(m^2 (98-91 {\cal A}(Q)+15 {\cal A}'(Q) (1-7 Q)) \
\phi ^2-9
(7 {\cal A}(Q)+{\cal A}'(Q)-7 {\cal A}'(Q) Q) V_{\text{eff}}\right)\right) \right\},
\label{alphaMWI}
\end{eqnarray}
and
\begin{eqnarray}
\beta_s &\simeq& 
\frac{m^6 M_{\text{Pl}}^6 }{343 Q^3
V_{\text{eff}}^6} \left\{ m^6 (-1134+810 {\cal A}(Q)+25 Q \left[32 {\cal A}'(Q)+5 {\cal A}''(Q) Q\right]) \phi^6
\right.
\nonumber \\
&-& \left.  2 m^4 (-812+646 {\cal A}(Q)+15 Q (68 {\cal A}'(Q)+15 {\cal A}''(Q) Q)) \phi ^4 V_{\text{eff}}
\right.
\nonumber \\
&+& \left. 4 m^2 (-28-10 {\cal A}(Q)+9 Q (26 {\cal A}'(Q)+15 {\cal A}''(Q)
Q)) \phi ^2 V_{\text{eff}}^2-216 (2 {\cal A}(Q)+Q (-2 {\cal A}'(Q)+{\cal A}''(Q) Q)) V_{\text{eff}}^3 \right\},
\nonumber \\
\label{betaMWI}
\end{eqnarray}
\end{widetext}
where $V_{\text{eff}}$ is given by Eq.~(\ref{pot}) and the function ${\cal A}(Q)$ is derived using the growth
function $G(Q)$ as defined in the next section and given by Eq.~(\ref{GQ}).

We solve the complete system of dynamical equations for the hybrid inflation model and compare the numerical
results with the analytical ones coming from the Eqs.~(\ref{nsMWI}), (\ref{alphaMWI}) and (\ref{betaMWI}).
As in Ref.~\cite{Berghaus:2019whh}, we consider the case of the model consisting of a pure $SU(3)$, 
with relativistic degrees of freedom $g_*=17$, consisting of two polarizations per eight gauge bosons plus one
for the axion. The gauge coupling is assumed to be $\alpha_g=0.1$ and $\kappa(\alpha_g, N_c,N_f)\alpha_g^5 =10^{-3}$
for the coefficient in the dissipation coefficient in Eq.~(\ref{upsilon}).
As also shown in Ref.~\cite{Berghaus:2019whh}, the strong dissipative regime ($Q\gg 1)$ in the model with the
hybrid type of potential Eq.~(\ref{Vhybrid}) is favored when the parameters $M$, $g$ and axion term $f$ satisfies
$g f/M \lesssim 10^{-8}$. We also recall that when assuming a QCD axion, astrophysical constraints~\cite{Sikivie:2020zpn} put
a lower  bound $f \gtrsim 10^9$ GeV on the QCD axion decay constant. In our numerical experiments we have checked
that both the strong dissipative regime and acceptable values for $f$ satisfying the astrophysical bounds can be
achieved by taking for example $M\gtrsim 2.5 \times 10^{17} \,g$ GeV and choosing appropriate values for $g$ and 
for the other constant, $\lambda$, appearing in the potential Eq.~(\ref{Vhybrid}). 

We have studied two representative cases and in the Tab.~\ref{tab01} we summarize  the relevant parameters and quantities
obtained from the numerical analysis. Note that for the parameters considered, we have
the temperature at the end of the WI slightly smaller than the axion decay constant, $f > T_{\rm end}$, which
indicates an axion symmetry breaking still in the inflationary regime and, hence, prior to the end of WI.
As an aside regarding the hybrid inflation model studied here, we note that the inflaton field excursions for
examples shown in Tab.~\ref{tab01} are all sub-Planckian, $|\Delta \phi| \sim 0.04 M_{\rm Pl}$. Hence, 
the hybrid WI model studied here satisfies the swampland distance conjecture, while it
marginally satisfies the (refined) de Sitter conjecture~\cite{Ooguri:2018wrx,Garg:2018reu}, by having $\epsilon_{V_*} \ll 1$,
but $\eta_V \gtrsim 1 $.

\begin{widetext}
\begin{center}
\begin{table}[!htb]
\begin{tabular}{c|c|c|c|c|c|c|c}
\hline \hline 
$g$ & $Q_*$  & $m/M_{\rm Pl}$ & $\epsilon_{V_*}$ & $\eta_{V_*}$ & $\phi_*/M_{\rm Pl}$ & $T_{\rm end}$ [GeV] & $f$ [GeV]
\\ \hline 
 & & & & & & & \\
$5\times 10^{-8}$ & 200 & $1.41\times 10^{-16}$  & $0.01 $  & 1.13 & 0.15 & $1.25\times 10^{9} $  & $4.75 \times 10^{9} $
\\ & & & & & & &
\\  \hline 
 & & & & & & & \\
$10^{-8}$ & 350  & $7.25\times 10^{-18}$ & 0.03 & 1.86 & $0.14$ & $2.77\times 10^{8} $  & $1.85 \times 10^{9} $
\\ & & & & & & &
\\ \hline \hline
\end{tabular}
\caption{Model parameters and relevant quantities obtained from the MWI model with 
the hybrid inflaton potential. In all cases we have considered  $M\gtrsim 2.5 \times 10^{17} \,g$ GeV and
$\lambda=10^{-2}$. The number of e-folds $N_*$ at Hubble radius crossing is found from Eq.~(\ref{N*}) and it
is $N_*\simeq 49$ for the cases shown here.}
\label{tab01}
\end{table}
\end{center}
\end{widetext}

The number of e-folds $N_*$ at Hubble radius crossing is found from the relation
obtained in WI (see, e.g., Ref.~\cite{Das:2020xmh} for details)
\begin{equation}
\frac{k_*}{a_0 H_0} = e^{-N_*} \left[ \frac{43}{11 g_s(T_{\rm end})}
  \right]^{1/3} \frac{T_0}{T_{\rm end}} \frac{H_*}{H_0},
\label{N*}
\end{equation}
where $T_{\rm end}$ is the temperature at the end of WI, $H_0$ is the Hubble parameter
today and for which  we assume the Planck result, $H_0=67.66\, {\rm km}\, s^{-1}
{\rm Mpc}^{-1}$ [from the Planck Collaboration~\cite{Aghanim:2018eyx},
  TT,TE,EE-lowE+lensing+BAO 68$\%$ limits,  $H_0 = (67.66 \pm 0.42)\,
  {\rm km}\, s^{-1} {\rm Mpc}^{-1}$], $T_0$ is the CMB
temperature today,  $T_0 = 2.725\, {\rm K}= 2.349
\times 10^{-13}\, {\rm GeV}$, while the for the pivot scale $k_*$ we
take the Planck value $k_* = 0.05/{\rm Mpc}$ and we also use the convention $a_0=1$.  
{}For $g_s(T_{\rm end})$ we assume the MWI value, $g_s(T_{\rm end}) =17$. 

\begin{widetext}
\begin{center}
\begin{table}[!htb]
\begin{tabular}{c|c|c|c|c|c|c}
\hline \hline $Q_*$ & $n_{s_*,{\rm analytical}}$ & $n_{s_*,{\rm numerical}}$&
$\alpha_{s_*,{\rm analytical}}$ & $\alpha_{s_*,{\rm numerical}}$ & $\beta_{s_*,{\rm analytical}}$&
$\beta_{s_*,{\rm numerical}}$ \\ \hline

 &  &  &  &&& \\ 
200 & $0.9611$ & $0.9612$ & $ -2.0 \times 10^{-4} $ & $ -1.9\times 10^{-4} $ & $-1.8 \times 10^{-6} $ & $-2.0\times 10^{-6}$\\ 
 &  &  & &&& \\ 
\hline

 &  &  &  &&&\\ 
350 & $0.9642$ & $0.9643$  & $-1.9\times 10^{-4}$ & $-1.9\times 10^{-4}$ & $-1.5\times 10^{-6}$ & $-1.1\times 10^{-6}$\\ 
 &  &  & &&& \\ 
\hline
\hline
\end{tabular}
\caption{Analytical estimation of $n_s$, $\alpha_s$ and $\beta_s$, along also with their
full numerical determination,  for the 2 cases considered in Tab.~\ref{tab01}.}
\label{tab02}
\end{table}
\end{center}
\end{widetext}

The numerical and analytical results obtained for
$n_s$, $\alpha_s$ and $\beta_s$ are given in Tab.~\ref{tab02}.
We note from the results from Tab.~\ref{tab02} that the analytical and numerical results
for $n_s$ and $\alpha_s$ agrees quite well. There are discrepancies between the values 
for $\beta_s$, specially for the lower dissipative case studied, $Q_*=200$, but that can be attributed 
to the lack of numerical precision
in the determination of $\beta_s$ in that case, recalling that to obtain $\beta_s$ numerically, it requires three derivatives 
of the scalar power spectrum with
respect to the number of e-folds. In any case, these results also corroborate the previous
ones using the approximated form of the power spectrum, with $\alpha_s$ and $\beta_s$ having
the same order of magnitudes as obtained previously, Eqs.~(\ref{alphas1}) and (\ref{betas1}),
respectively. We have not quoted the results for the tensor-to-scalar ratio $r$, but as typical for WI in the strong
dissipative regime, $r$ is too small to be accessible by any future observation. {}For the dissipation values
quoted in Tab.~\ref{tab01}, $r\lesssim 10^{-25}$.

We now study another type of WI models that can allow for large
dissipation and are still consistent with the Planck observations.
The objective is to check whether with these models we can break the
large hierarchy between $\alpha_s$ and $\beta_s$.
 
\section{Running and running of the running of the scalar spectral index in MWI with exponential potentials}
\label{MWI-expo}

In CI, steep exponential potentials of the form 
\begin{eqnarray}
V(\phi)=V_0e^{-\alpha\phi/M_{\rm Pl}},
\end{eqnarray}
leads to power-law type inflation~\cite{Liddle:1988tb}, where
inflation does not exit gracefully in standard general
relativity. However, it was shown in Ref.~\cite{Das:2020lut} that WI
with dissipative coefficient of the form $\Upsilon(T)\propto T^p$,
with $p>2$, can gracefully exit inflation with such an exponential
potential. The MWI model~\cite{Berghaus:2019whh} is, thus, such a
model where the dissipative coefficient scales with the cubic power of
the temperature of the thermal bath, $\Upsilon(T)\propto T^3$.

A MWI model with such an exponential potential was studied in
Ref.~\cite{Das:2019acf}.\footnote{Recently such exponential potential is also studied in the context of power-law warm inflation \cite{Alhallak:2022szt}.} Though inflation ends gracefully in such a
model, it was shown there that such a combination yields way too large
red-tilt in the scalar spectral index to be compliant with the current
observation. Therefore, such models are not viable models of
inflation. 

A generalized form of the exponential potential, given as
\begin{eqnarray}
V(\phi)=V_0e^{-\alpha(\phi/M_{\rm Pl})^n}, 
\label{potexp}
\end{eqnarray}
with $n>1$, was then studied in Ref.~\cite{Das:2020xmh}, which shows
that this model not only exits inflation gracefully, but also fully
satisfies all the observational constraints (comply with the observed
values of scalar spectral index $n_s$ and the tensor-to-scalar ratio
$r$) as well as to be in tune with the Swampland conjectures (see,
e.g., Table~1 of Ref.~\cite{Das:2020xmh}). This same model was also 
considered in the context of WI for quintessential inflation~\cite{Lima:2019yyv},
displaying interesting features as far as the late Universe physics is
concerned.
Therefore, here we will
further analyze such a model to determine the running and running of
the running of the scalar spectral index that are yielded by this
model. 

In the model that we are presently considering, the explicit form for
the function $G(Q_*)$ appearing in the scalar of curvature power
spectrum in WI, Eq.~(\ref{full-power}), has been derived in
Ref.~\cite{Das:2020xmh} and it is given as\footnote{See Appendix~B of
  Ref.~\cite{Das:2020xmh} for the full derivation of the scalar power
  spectrum in this model.}
\begin{eqnarray}
  G(Q_*)&=&\frac{1+6.12Q_*^{2.73}}{(1+6.96Q_*^{0.78})^{0.72}}
  \nonumber\\ &&+\frac{0.01Q_*^{4.61}(1+4.82
\times 10^{-6}Q_*^{3.12})}{(1+6.83\times10^{-13}Q_*^{4.12})^2}.
\label{GQ}
\end{eqnarray}

We will first derive the scalar spectral index $n_s$, its running
$\alpha_s$ and running of its running  $\beta_s$ analytically in this
model. Note that, although the values of $n_s$ in this model has been
reported previously in Table~1 of Ref.~\cite{Das:2020xmh}, they have
been obtained numerically. 
To the best of our knowledge, this is for
the first time that we aim to determine both $n_s$, $\alpha_s$ and
$\beta_s$ in any WI model analytically. 

If we ignore the thermal distribution factor $(1+2n_*)$ of the
inflaton field in Eq.~(\ref{Fk}), which can be justified when working
in the strong dissipative regime $Q\gg1$ (see, e.g.,
Ref.~\cite{Ramos:2013nsa}), then the scalar power spectrum simplifies
to the form 
\begin{eqnarray}
\Delta_{\mathcal
  R}(k)=\frac{\sqrt3}{4\pi}\frac{H^3T}{\dot\phi^2}\frac{Q}{\sqrt{3+4\pi
    Q}}G(Q).
\label{scalar-ps}
\end{eqnarray}
This is the form of the power spectrum that we will use to calculate
the spectral index and its running and running of its running
analytically. Such an approximation of the power spectrum will later
be justified by obtaining $n_s$, $\alpha_s$ and $\beta_s$ numerically
and matching them with the analytically obtained results. 

{}First, we calculate the scalar spectral index $n_s$.  Using the
scalar spectrum given in Eq.~(\ref{scalar-ps}) we get 
\begin{eqnarray}
n_s-1&=&3\frac{d\ln H}{dN}+\frac{d\ln
  T}{dN}-2\frac{d\ln\dot\phi}{dN}\nonumber\\ &&+\left(\frac{3+2\pi
  Q}{3+4\pi Q}+Q\frac{d\ln G(Q)}{dQ}\right)\frac{d\ln Q}{dN}.
\end{eqnarray}
{}From the equations in Eq.~(\ref{equations}), after some straightforward algebra, we get
the scalar spectral index as 
\begin{eqnarray}
n_s\approx1-\frac{(14-10{\mathcal A}(Q))\epsilon_V+6{\mathcal
    A}(Q)\eta_V}{7Q},
\label{spectral-index2}
\end{eqnarray}
where $\mathcal{A}(Q)$ has been defined in Eq.~(\ref{AQ}).

We now analytically determine $\alpha_s$. Using the form of $n_s$ from
Eq.~(\ref{spectral-index2}), we obtain
\begin{eqnarray}
\alpha_s&=&\frac{1}{7Q}\left\{-(14-10{\mathcal
  A}(Q))\frac{d\epsilon_V}{dN}-6{\mathcal
  A}(Q)\frac{d\eta_V}{dN}\right.\nonumber\\ &&+\left[(14-10{\mathcal
    A}(Q))\epsilon_V+6{\mathcal
    A}(Q)\eta_V\right.\nonumber\\ &&\left.\left.+(10\epsilon_V-6\eta_V)Q
    {\mathcal
      A}'(Q) \right]\frac{d\ln Q}{dN} \right\}.
\end{eqnarray}
Using now that
\begin{eqnarray}
&&\frac{d\epsilon_V}{dN}=\frac{4\epsilon_V^2-2\epsilon_V\eta_V}{Q},\\ &&
  \frac{d\eta_V}{dN}=\frac{2\epsilon_V\eta_V-\xi_V^2}{Q},
\end{eqnarray}
the expression for the running of the scalar index can be computed to
be
\begin{eqnarray}
\alpha_s&=&\frac{1}{49Q^2}\left[4\left(-63+45{\mathcal
    A}(Q)+25Q{\mathcal A}'(Q)\right)\epsilon_V^2 \right.
  \nonumber \\ &-& \left. 8\left(-14+13{\mathcal
    A}(Q)+15Q{\mathcal
    A}'(Q)\right)\epsilon_V\eta_V\right.\nonumber\\ &&\left.+36\left(-{\mathcal
    A}(Q)+Q{\mathcal
    A}'(Q)\right)\eta_V^2+42\mathcal{A}(Q)\xi_V^2 \right].
\nonumber \\
\label{running}
\end{eqnarray}

To determine $\beta_s$, one requires
\begin{eqnarray}
\frac{d\xi^2_V}{dN}=\frac1Q[(4\epsilon_V-\eta_V)\xi^2_V-\omega_V^3].
\end{eqnarray}

Using the form of $\alpha_s$ from Eq.~(\ref{running}), $\beta_s$ can
be written as 
\begin{widetext}
\begin{eqnarray}
\beta_s=\frac{1}{343 Q^2}&&\left[8\left(-1764 + 1260 {\mathcal A}(Q) +
  1050 Q {\mathcal A}'(Q)+ 125 Q^2{\mathcal
      A}''(Q)\right)\epsilon_V^3\right.\nonumber\\ &&-24\left(-490
  + 392 {\mathcal A}(Q) + 490  Q {\mathcal A}'(Q)+ 75
  Q^2{\mathcal
      A}''(Q)\right)\epsilon_V^2\eta_V\nonumber\\ &&+8 \left(-196
  + 56  {\mathcal A}(Q)+ 504 Q {\mathcal A}'(Q)+ 135
  Q^2{\mathcal
      A}''(Q)\right)\epsilon_V\eta_V^2\nonumber\\ &&+28 \left(-28
  + 68{\mathcal A}(Q)+ 45 Q{\mathcal
      A}'(Q)\right)\epsilon_V\xi_V^2-216 Q^2{\mathcal
      A}''(Q)\eta_V^3\nonumber\\ &&\left.+42 \left(5 {\mathcal
    A}(Q) - 18  Q{\mathcal A}'(Q)\right)\eta_V\xi_V^2-294
  {\mathcal
    A}(Q)\omega_V^3\right]-\frac{2(10\epsilon_V-6\eta_V)}{7Q}\alpha_s.
\end{eqnarray}
\end{widetext}

The equations for $n_s$, $\alpha_s$ and $\beta_s$ can also be expressed explicitly in terms of the
inflaton field $\phi$ using the potential Eq.~(\ref{potexp}), but that only complicates
more the form of the expressions and we do not require to perform that explicitly here.
In the Tab.~\ref{tab1}, we will furnish explicitly the numerical values for the required
slow-roll parameters.

\begin{widetext}
\begin{center}
\begin{table}[!htb]
\begin{tabular}{c|c|c|c|c|c|c|c|c}
\hline \hline Model & $V_0$ (GeV$^4$) & $V_*^{1/4}/M_{\rm Pl}$ &
$\epsilon_{V_*}$ & $\eta_{V_*}$ & $\xi_{V_*}^2$ & $\omega^3_{V_*}$ & $T_{\rm end}$ [GeV] & $f$ [GeV]
\\  \hline $n=2$ &  &  &  &  &  &&  \\  $\alpha= 9.6$& $3.07\times
10^{38}$ & $1.14\times 10^{-9}$ & 31.7 & 44.2 & 367.72 & $-1.38 \times
10^5$ &  $2.1 \times 10^7$ &   $7.3 \times 10^9$
\\  $Q_*=850.96$ &  &  &  &  & &&  
\\  \hline 
$n=3$ &  &  &  &  & &&
\\  $\alpha= 2.5$& $1.82\times 10^{39}$ & $9.55\times 10^{-10}$ & 25.9
& 37.1 & 507.73 & $-4.16 \times 10^4$   & $3.4\times 10^7$ &  $8.8\times 10^9$
\\  $Q_*=740.15$ &  &  &  &  &  &&
\\  \hline $n=4$ &  &  &  &  &  &&  
\\  $\alpha= 0.45$& $3.59\times
10^{39}$ & $1.34\times 10^{-9}$ & 25.0 & 36.56 & 603.87 & $-2.6 \times 10^4$  & $3.9\times 10^7$ & $9.2\times 10^9$
\\  $Q_*=719.68$ &  &  &  &  &  &&  
\\  \hline $n=5$ &  &  &  &  & &&
\\  $\alpha= 0.06$& $6.01\times 10^{39}$ & $1.69\times 10^{-9}$ & 24.1
& 35.5 & 612.63 & $-1.89 \times 10^4$  & $4.3\times 10^7$  &  $9.6\times 10^9$
\\  $Q_*=699.53$ &  &  &  &  &  &&
\\  \hline \hline
\end{tabular}
\caption{Values for the parameters considered in the analytical estimation
  of $n_s$, $\alpha_s$ and $\beta_s$ for 4 sets of the chosen model parameters. The number of e-folds $N_*$ is obtained
from Eq.~(\ref{N*}) and it is $N_*\simeq 48$ for all the cases shown here~\cite{Das:2020xmh}. }
\label{tab1}
\end{table}
\end{center}
\end{widetext}

We now estimate the values of $n_s$, $\alpha_s$ and $\beta_s$ using
the analytical forms of these quantities that we derived above. To
obtain these values, we need to fix $n$, $\alpha$ and $Q_*$,  the
values of which have been taken from Table~1 of
Ref.~\cite{Das:2020xmh} which produce $n_s$ and $r$ that are in tune
with the observations\footnote{Note that the values of
  $V_*^{1/4}/M_{\rm Pl}$ have been wrongly quoted in Table~1 of
  \cite{Das:2020xmh}. {}Furthermore, in the same reference we have incorrectly
quoted the estimate for the axion decay constant for this model, due
to the use of an underestimated value for the gauge field coupling constant $\alpha_g$ used
in the numerical code in that work. We have now corrected both of these values in the present work.}.
In the table we have also quoted the temperature at the end of WI, $T_{\rm end}$ and
the value obtained for the axion decay constant. To obtain $f$, we made the same
considerations regarding the gauge sector coupled to the scalar field as in the
previous section for the MWI model with the hybrid potential.
We have considered QCD like values for the parameters, with relativistic degrees of freedom $g_*=17$, 
the gauge coupling is assumed to be $\alpha_g=0.1$ and $\kappa(\alpha_g, N_c,N_f)\alpha_g^5 =10^{-3}$
for the coefficient in the dissipation coefficient in Eq.~(\ref{upsilon}).
Note that in all the cases considered here, we have that $T_{\rm end} < f$, pointing to
cases that the axion is already in a symmetry broken phase at the end of WI.

In the Table~\ref{tab1}, we give the values of the required slow-roll
parameters required in the derivation of $n_s$, $\alpha_s$ and
$\beta_s$. We give also the normalization $V_0$ of the potential for
each of the models and the scale of the inflaton potential at Hubble
radius crossing, $V_*$.  The analytically estimated values for $n_s$,
$\alpha_s$ and $\beta_s$ that are obtained for each of those models
are then shown in Table~\ref{tab2}. In the same table it is also given
the numerically obtained vales for $n_s$, $\alpha_s$ and $\beta_s$,
where the full scalar power spectrum, given in Eq.~(\ref{full-power}),
have been used to determine them.

\begin{widetext}
\begin{center}
\begin{table}[!htb]
\begin{tabular}{c|c|c|c|c|c|c}
\hline \hline Model & $n_{s_*,{\rm analytical}}$ & $n_{s_*,{\rm numerical}}$&
$\alpha_{s_*,{\rm analytical}}$ & $\alpha_{s_*,{\rm numerical}}$ & $\beta_{s_*,{\rm analytical}}$&
$\beta_{s_*,{\rm numerical}}$ \\ \hline

$n=2$ &  &  &  &&& \\ 
$\alpha= 9.6$ & $0.9651$ & $0.9648$ &
$-5.9\times 10^{-3}$ &
$-6.1\times 10^{-3}$ & $-8.6\times 10^{-6}$ & $-2.7\times 10^{-5}$\\ 
$Q_*=850.96$ &  &  & &&& \\ 
\hline

$n=3$ &  &  &  &&&\\ 
$\alpha= 2.5$& $0.9697$ & $0.9689$  &
$-4.2\times 10^{-3}$ &
$-4.4\times 10^{-3}$ & $-1.2\times 10^{-4}$ & $-1.3\times 10^{-4}$\\ 
$Q_*=740.15$ &  &  & &&& \\ 
\hline

$n=4$ &  &  &  &&&\\ 
$\alpha= 0.45$ & $0.9662$ & $0.9655$ &
$-3.7\times 10^{-3}$ &
$-3.8\times 10^{-3}$ & $-1.3\times 10^{-4}$ & $-1.4\times 10^{-4}$\\ 
$Q_*=719.68$ &  &  &  &&& \\ 
\hline

$n=5$ &  &  &  &&& \\ 
$\alpha= 0.06$ & $0.9654$ & $0.9645$ &
$-3.4\times 10^{-3}$ &
$-3.5\times 10^{-3}$ & $-1.3\times 10^{-4}$ & $-1.4\times 10^{-4}$\\ 
$Q_*=699.53$ &  &  &  &&& \\ 
\hline
\hline
\end{tabular}
\caption{Analytical estimation of $n_s$, $\alpha_s$ and $\beta_s$, along also with their
full numerical determination,  for 4 sets of chosen model.}
\label{tab2}
\end{table}
\end{center}
\end{widetext}

{}From the results shown in Table~\ref{tab2}, we notice that the
analytical results match well with the numerical results. Therefore,
the approximation we made to the scalar power spectrum in
Eq.~(\ref{scalar-ps}) is justified. We also note that MWI model with
generalized exponential potentials also produces negative $\alpha_s$
and $\beta_s$. However, in comparison to the MWI model with hybrid
potential, it produces one order higher $\alpha_s$ and two-order
higher $\beta_s$. These results are promising in the sense that
indicates that the large hierarchy observed between $\alpha_s$ and
$\beta_s$, which appears in the vanilla CI models, can be broken in
the context of WI.

\section{Discussion and Conclusion}
\label{conclusion}

There is a hint in the final data released by {\it Planck} that the
running and the running of the running of the scalar spectral index
might be larger than predicted in standard cold inflation models, with
$\alpha_s$ of the order of $10^{-2}$, and the running of the running
might be larger than the running itself~\cite{Planck:2018jri}. Vanilla
cold inflationary models preferred by {\it Planck} data fails to
produce such large (and also positive) running and running of the
running~\cite{Escudero:2015wba, vandeBruck:2016rfv}.  Thus, it is of
importance to seek for inflationary models which can follow these
trends for the runnings hinted by the {\it Planck} data.

In this paper, we analyzed a variant inflationary scenario, namely
warm inflation, in a strong dissipative regime ($Q\gg1$). Previously,
the running and the running of the running of the scalar spectral
index was studied in Ref.~\cite{Benetti:2016jhf}, where WI was
realized in weak dissipative regime $(Q\ll1)$.  To analyze WI in the
strong dissipative regime, we chose a particular model of WI, namely
the minimal warm inflation model first studied in
Ref.~\cite{Berghaus:2019whh}. In this model, the dissipative term is
proportional to the cubic power of the temperature of the thermal bath
($\Upsilon\propto T^3$). MWI was first studied with a hybrid potential
in Ref.~\cite{Berghaus:2019whh} to obtain a red-tilted scalar
spectrum. However, later in Ref.~\cite{Das:2020xmh}, MWI was studied
with generalized forms of the exponential potential and it was shown
in that reference that not only this model is in accordance with the
current observations, but can also accommodate the Swampland
conjectures.  We determine $\alpha_s$ and $\beta_s$ in both of these
MWI models. Explicit analytical expressions for the running and for
the running of the running were derived in the present paper. To the
best of our knowledge, this is for the first time that the running and
the running of the running of scalar spectral index in the context of
WI have been analyzed analytically, in special in the strong
dissipative regime of WI. We believe that this work will motivate
further exploration of other WI models along the same line to see
whether WI can produce the {\it Planck}--hinted running and running of
the running. This is of particular importance, given that the next
generation of cosmological observations might further improve on the
determination for the values for these cosmological parameters.

\begin{acknowledgments}

R.O.R. acknowledges financial support of the Coordena\c{c}\~ao de
Aperfei\c{c}oamento de Pessoal de N\'{\i}vel Superior (CAPES) -
Finance Code 001 and by research grants from Conselho Nacional de
Desenvolvimento Cient\'{\i}fico e Tecnol\'ogico (CNPq), Grant
No. 307286/2021-5, and from Funda\c{c}\~ao Carlos Chagas Filho de
Amparo \`a Pesquisa do Estado do Rio de Janeiro (FAPERJ), Grant
No. E-26/201.150/2021. 

\end{acknowledgments}



\end{document}